\documentclass[runningheads,a4paper]{llncs}

\usepackage{amssymb}
\usepackage{graphicx}

\usepackage{url}
\urldef{\mailsa}\path| yayank@usu.ac.id, mahyunst@yahoo.com |
\newcommand{\keywords}[1]{\par\addvspace\baselineskip
\noindent\keywordname\enspace\ignorespaces#1}

\begin{document}

\mainmatter  

\title{Generating Strategic IS: Towards the Winning Strategy}

\titlerunning{Generating Strategic IS: Towards the Winning Strategy}

\author{Maria Elfida$^{1}$%
\thanks{}%
\and Mahyuddin K. M. Nasution$^{1,2}$}

\authorrunning{M. Elfida dan M. K. M. Nasution}

\institute{$^{1}$Program S2 Teknik Informatika, FASILKOM-TI,\\
Universitas Sumatera Utara, Padang Bulan 20155 USU, Medan, Indonesia.\\
$^{2}$Departemen Teknologi Informasi, FASILKOM-TI,\\
Universitas Sumatera Utara, Padang Bulan 20155 USU, Medan, Indonesia.\\
\mailsa\\}

\toctitle{draf}
\tocauthor{}
\maketitle

\begin{abstract}
In modern era, the role of information system in organization has been taken many discussions. The models of information system are constantly updated. However, most of them can not face the changing world. This paper discusses an approach to generating of strategic information system based on features in organization. We proposed an approach by using disadvantages in some tools of analysis whereby the lack of analysis appear as behaviour of relation between organisation and the world.
\keywords{Information system, information technology, feature, organization, knowledge}
\end{abstract}

\section{Introduction}

The success of information system (IS) has been one of themes in many discussions in the past two decades \cite{delone1992,delone2003}. The rapid growth in the use of information systems has led to several changes in the workflow of both the private and public sectors, mainly about providing the units for working, servicing, implementing all works in environment of information technology \cite{feindt2002}. A change is to compete with one another. The changes are as attempts to develop efficiency services, improved products, and better systems. The change for other changes, that to be the implementation of and use for what do the information technology is constantly changing. An information system is always by means of the concept, planning, design, implementation, and use of advanced technology either from the beginning or the capacity for self-upgrading and self-updating. Why are so many implementations and information systems created and so few used? Should information system be something strategic, so how? 

A strategic information system is an information system for helping to overcome the problem of inefficient public service and information delivery in the public sector \cite{melville2004} by using the information technology effectively. The information system thus has been identified as vital in order to avail continually be available to generate the services. The Information system is critical in this age whereby the knowledge has been a part of ubiquitous of environments and their strategies, the knowledge as resources of competence. However, there is little attention about the study of information system as strategic factor to an environment such as enterprise, organization, or special in Indonesia. We assure that the competence will become an important part of information system, and to support the achievement towards "the winning strategy". This paper aims to address an approach to identify strategic information system for an organization in a changing world.

\section{Related Work}

Information system is one of important studies in information area since its universal model to be important consequences for many organization and enterprise \cite{delone1992,delone2003}, and for the next period received many improvements \cite{galletta1989} based on some factors. Some of factors are the management style, size, goals, vision, mission and culture. However, the researchers also have identified organisational structure, organisational size, managerial information technology knowledge, top management support, financial resources, goal alignment and budgeting method as factors that affect the success of information systems.
Currently, the information technology has supported many applications for understanding the behavior people, groups of people or organizations \cite{nasution2010,nasution2011}. The information technology supports and influences business strategies and objectives \cite{reyes2005}. In general, in many books about analysis and design system, they explained that to implement and settle an information system becomes better and more better, constantly used the system development life cycle: planning, analysis, design, implementation, and testing. Therefore, the success of information system depends on the planning mainly. The process of planning will determine the success of implementation and using the information system in an organization \cite{mentzas1997}. Planning for information system requires an organization to consider the current situation and potential future, and to select most appropriate way to achieve objectives. In analysis, any organization needs to conduct the critical success factor (CSF) \cite{jockart1979}for ensuring successful competitive performance will be achieve, to make the process analysis (PA) technique \cite{ward2002} for understanding the business process that are supporting objectives, to implement SWOT (strengths, weaknesses, opportunities, and threats) analysis for modeling a situation for organization \cite{robson1994}, to do the normative analysis (NA) \cite{davis1982} for finding the limitations of organization, to study basically the system by end-means analysis (EMA) for emphasizing the identification of reliable things in organization, to conduct business strategy analysis (BSA) for deriving the essential relations between organization and business, to analysis with the value chain analysis (VCA) for looking for the opportunities that can be exploited or supported by information technology \cite{porter1980}, and it needs to model the dynamic world with Porter's five forces (PFF) model for identifying the opportunities in business. However, features of organization on planning sub-instances cannot be expressed clearly. Note: A planing as the gray (from dark to light or light to dark) is as the contribution levels. Therefore, we proposed an approach to address the changing an information system becomes strategic information system. 

\section{The Proposed Approach for Generating Strategic Information System}

Each organization as a system exists in an environment, i.e., a world that consists of classes of knowledge \cite{nasution}. Therefore, an organization has something in isolation and also things in sharing. An organization starts with idea, concept, vision and mission, whereby the organization achieves its objectives, but an organization is always in a predictable path, going on achievements and performances. A world can be considered a space of event whereby an organization will connect to any event, each event can measure in either discrete or continue. An event can be recognized clearly as a number, a count, or a value, but sometimes it can not identified. An event sometimes depends on another event. The both events have any relation. Let us model an organization as a set of entities, then entities exists in both an organization and a world, and therefore each entity has the stable attributes and the flexible attributes. Thus, the event can be considered as uncertainty for some situations and can be measured by the probability \cite{nasution2012}. Therefore, we group the features of information system into six categories as the beginning of analysis, i.e. an approach that classifies the features into internal, external, quantitative, qualitative, history, and forecasting categories, where each feature is potentially same opposition to core, to generate a strategic information system, based on a comprehensive view, scope of data/information, mechanisms, definiting requirements, high level goal and knowledge-oriented paradigm.

Some features internally in organisation as the advantage and if we confront them to organisation's environment will be the complex situation. Size of organization become bigger if the organization was expanded area of its colonies, the branches is needed to represent organization, and then information system is one panacea to control all resources in organization, but when the organization take a system to supervise all, the organization increasingly exposed to the world and it is most difficult to control them. In strategic information system, the system needs to define the internal features of organization as a basic modal in information system for facing an environment and also the system needs to identify the external features from outside as input for defending own self from an attack. As a strategy, the organization must know that the internal and the external features have relationship, and the relation between them always hidden in background of organization or the features. Each feature has the latent threat in relations. Therefore, each strategic information system requires the support in multi level analysis, the information/data modelling, and the background of knowledge in having sound theoretical basis. In this case, a comprehensive view of organization requirements will capture requirements of information system. 

Some features exist as information quantitative, something easily we recognize it. In strategic information system, it is represented as statistics, graphics or other infrastructures of multimedia. However, other features we can identify only through stable attributes of another feature in their relations. An information model should take an embraced view to conceptualize the requirements from data to knowledge in order to an environment can be determine existing organization uniquely with information system. There are many features in qualitative condition and this is most subjective, because the values of them depend on the exercise of members in organization or they are outside. Therefore, requirements about data/information must cover in wider scope. 

The important feature comes and goes based on time in a timeline of organization. The robust feature always exist, and it as a core in information system. The core of feature is a benchmark for evaluating the information system while other features are the constraints for supporting the forecasting. Some tools in statistic and optimization are the best assistants to measure the importance of features and their relations. The strategic information system must enable to generate the temporary situations of organization and predictable some the choices as solution. In the system therefore there is a mechanism to address the complex situation of an organization whereby the organization regenerate high level goal for learning to the strategic information system.

The features are phenomena about organization and information system is a paradigm for understanding the phenomena, but in some conditions not each paradigm enable we use to understand any phenomena, there is a bridge the gap always exists between them. Anyway, everything has limitations, only human with their knowledge has a container of the strategy making. A strategic information system therefore must support knowledge-oriented paradigm, i.e. a system based on knowledge in overall of system as a winning strategy.

\section{Conclusion and Future Work}

The strategic information system can be generated based on an approach with six advantages of data as basic modal of organization to face the change world. The approach implicitly covers all attributes and values of features to take the implications for getting the winning strategy. Our near future work is to develop a framework to generate the background of knowledge about strategic information system.

\end{document}